\documentclass{llncs}
\usepackage{graphicx}
\usepackage{amsmath}

\newcommand{\de}{\partial}

\renewcommand{\b}[1]{\boldsymbol{#1}}
\begin{document}

\mainmatter

\title{Chaos in a simple cellular automata model of a uniform society}
\author{	
	Franco Bagnoli\inst{1,2,4,5} \and  
	Fabio Franci\inst{2,4} \and 
	Ra\'ul Rechtman\inst{3}
}

\institute{
	Dipartimento di Energetica, 
	Universit\`a di  Firenze, \\ Via S. Marta, 3 
	I-50139 Firenze, Italy.
	\email{bagnoli@dma.unifi.it} \\
	\and
	Centro Interdipartimentale per lo Studio delle Dinamiche Complesse,\\ 
	Universit\`a di  Firenze, Italy.
	\email{fabio@dma.unifi.it} 
 \and
  Centro de Investigac\'\i{}on en Energ\'\i{}a, 
 UNAM, \\62580 Temixco, Morelos, Mexico. 
 \email{rrs@teotleco.cie.unam.mx}
 \and
 INFM, Sezione di Firenze
 \and
 INFN, Sezione di Firenze.
}

\maketitle
\setcounter{footnote}{0}

\begin{abstract}
In this work we study the collective behaviors arising in a model of a simplified homogeneous society. Each agent is modeled as a binary ``perceptron'', receiving  neighbors' opinions as inputs and outputting one of two possible opinions, according to the conformist attitude and to the external pressure of mass media. For a neighborhood size greater than three, the system shows a very complex phase diagram, including a disordered phase and chaotic behavior. We present analytic calculations, mean fields approximation and numeric simulations for different values of the parameters. 
\end{abstract}

\section{Introduction}\label{introduction}

In recent years there has been a great interest in the applications of physical approaches to a quantitative description of social and economic processes \cite{haken,lorenz}. 
The development of the interdisciplinary field of the ``science of complexity'' has lead to the insight that complex dynamic processes may also result from simple interactions, and even social structure formation could be well described within a mathematical approach.

In this paper we are interested in modeling the human society in order to understand the process of opinion formation applied to political elections. Many different approaches have been taken by scientists 
in this matter. The simplest choice is to consider a uniform society, where all the persons are considered equal and have the same influence on all the others.
The second choice is to consider a society divided into two groups, a group of normal people  and a group of leaders (see Ref.~\cite{leader}).

Another possibility is to consider all the individuals different from one another. This can be the case of the Kauffman model, where every individual has different (randomly generated) features and interacts randomly with a certain number of other individuals. Despite its simplicity, this model is able to exhibit very complex behaviors, including chaos.

In this paper we start exploring the behavior of a model of an uniform society. By uniform we mean that everybody is exactly like the others, except for the expressed opinion (that may vary with time). All other parameters are exactly the same. 

We use a simple neural network model to approximate the process of opinion formation taking place in a society. With this approximation, the individual is represented by an automaton that receives some external inputs, elaborates them using a certain function, and elaborates a response. 

As a first step, we study the behavior of the simplest,  one-dimensional neural network, composed by binary perceptrons. 
Such a model has been successfully applied to anticipate personal preferences on products (see Ref.~\cite{BBF}).
The inputs for each perceptron are given by the opinions expressed by other persons in a local community, whose size is $R$. 

The parameters of the model are the influence of the mass-media ($H$), the weight $J$ assigned to the the local community (that can be thought as the result of education), and the conformist threshold $Q$. This last parameter models the empirical fact that even a strong mass-media campaign or a strong anti-conformist attitude cannot modify an opinion if it is supported by a strong majority in the local community. 
If the local majority is outside the thresholded intervals, the evolution rule is that of a Monte-Carlo simulation of an equilibrium system with heat-bath dynamics.

The system may  be trapped into one of the two absorbing states (uniform opinion), or exhibit several kinds of irregular behavior. Technically, it is an extension of a probabilistic cellular automata 
with a very rich phase diagram~\cite{Bagnoli_2001}, whose mean-field approximation was described in Ref.~\cite{acri02}. A detailed investigation of its behavior will be illustrated elsewhere~\cite{prl}.

In Sec.~\ref{Sec:model} we present this probabilistic cellular automata model and present the main results. In Sec.~\ref{Sec:lyap} we extend the concept of Lyapunov exponents to probabilistic cellular automata and present the results obtained for our model.
Finally, in Sec.~\ref{Sec:concl}, we summarize our work and draw some conclusions.

\section{The model}
\label{Sec:model}

Let $x_i^t$ be the opinion assumed by individual $i$ at time $t$. As usual in spin models, only two different opinions, -1 and 1, are possible. The system is made up of $L$ individuals arranged on a one-dimensional lattice with periodic boundary conditions. Time is considered discrete (corresponding, for instance, to election events).
The state of the whole system at time $t$ is given by $\boldsymbol{x}^t=(x_0^t,\dots,x_{L-1}^t)$ with
$x_i^t\in \{-1,1\}$.

The individual opinion is formed according to a local community ``pressure'' and a global influence. In order to avoid a tie, the local community is made up of $R=2r+1$ individuals, including the opinion of the individual himself at previous time. The average opinion of the local community around site (person) $i$ at time $t$ is denoted by $m=m^t_i = \sum_{j=- r}^{r} x^t_{i+j}$.

Let $J$ be a parameter controlling the influence of the local field in the opinion formation process and  $H$ be the external social pressure. 
One could think of $H$ as the television influence, and $J$ as the educational effects. The ``field'' $H$ pushes toward one opinion or the other, and people educated toward conformist will have $J>0$, while non-conformists will have $J<0$. 

The hypothesis of 
alignment with an overwhelming local majority is represented by the parameter $Q$, indicating the critical size of local majority. If $m<2Q-R$, then $x_i^{t+1}=-1$,
and if $m>R-2Q$, then $x_i^{t+1}=1$.

For simpler numerical implementation and plotting, the opinions ($-1$,$1$) are replaced by ($0$,$1$). Let us denote by $S$ the sum of opinions in the local community using the new coding. The ``magnetization'' can by expressed as
$m=2S-R$, and the probability $P_S=P(1|S)$ of expressing opinion $1$ given $S$ neighbors with opinion $1$ is:

\begin{equation}
\label{Eq:p_s}
P_S=\begin{cases}
 0 & \text{if $S<Q$;}\\
\dfrac{1}{1+\exp(-2(H+J(2S-R)))} & \text{if $Q \le S \le R-Q$;}\\
 1 & \text{if $S>R-Q$.}
 \end{cases}
\end{equation}

For $Q=0$ the model reduces to an Ising spin system. 
For all $Q>0$ we have two absorbing homogeneous states,
$\boldsymbol{x}=\boldsymbol{0}$  and 
 $\boldsymbol{x}=\boldsymbol{1}$,
corresponding to an infinite plaquette coupling in the statistical mechanical sense. 
With these assumptions, the model reduces to 
a one-dimensional, 
one-com\-ponent, totalistic cellular automaton with two absorbing
states.

The order parameter is the
fraction $c$ of people sharing opinion $1$\footnote{The usual order parameter for magnetic system is the magnetization $M = 2c-1$.}. 
It is zero or one  
in the two absorbing states, and assumes
other values in the active phase. The model is symmetric since the  two absorbing states have the same importance.


The quantities $H$ and $J$ range from $-\infty$ to $\infty$. 
For easy  plotting, we use the parameters 
$j = [1+\exp(-2J)]^{-1}$ and $h = [1+\exp(-2H)]^{-1}$ 
as control parameters, mapping the real axis ($-\infty, \infty$) to
the interval $[0,1]$. 

The fraction $c$ of ones in a
configuration and the concentration of clusters $\rho$ are defined by
\[
  c=\dfrac{1}{L}\sum_i S_i\quad\text{and}\quad%
 \rho=\dfrac{1}{L}\sum_i |S_i - S_{i+1}|.   
\]
In the mean-field approximation, the order parameters $c$ and $\rho$ are
related by $\rho=2c(1-c)$.
Both the uniform zero-state and one-state correspond to
$\rho\rightarrow 0$ in the thermodynamic limit. 

\begin{figure}
\begin{tabular}{ccc}
\includegraphics[width=6cm]{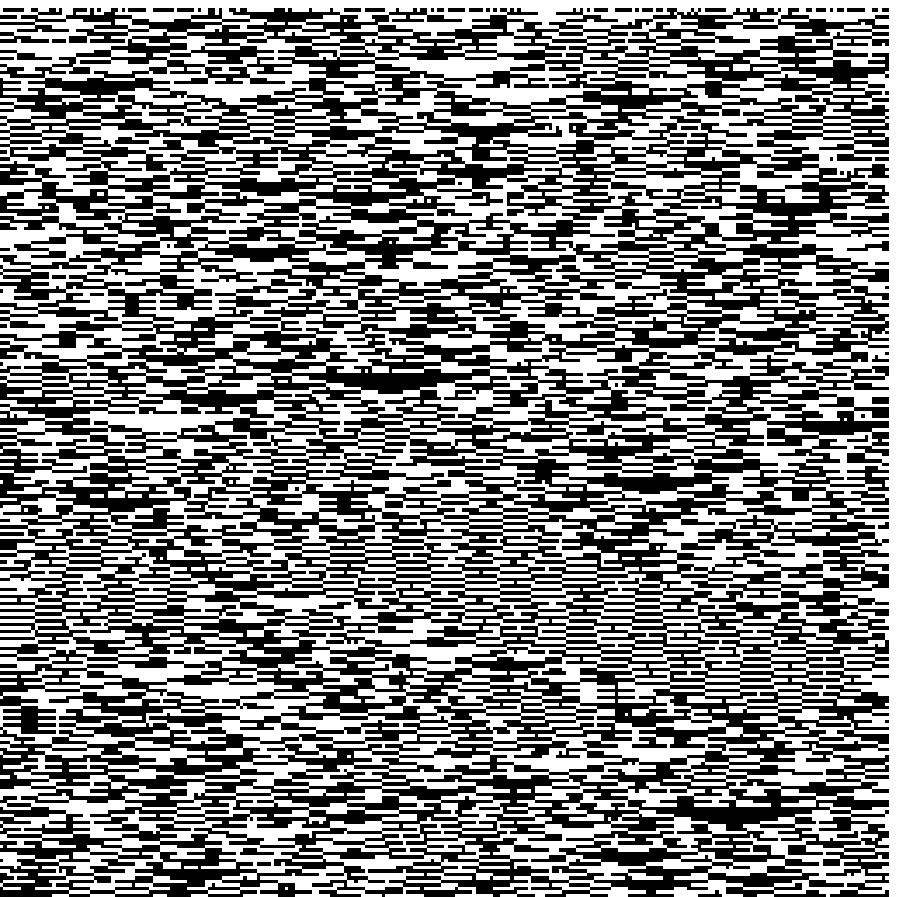} &
\includegraphics[width=6cm]{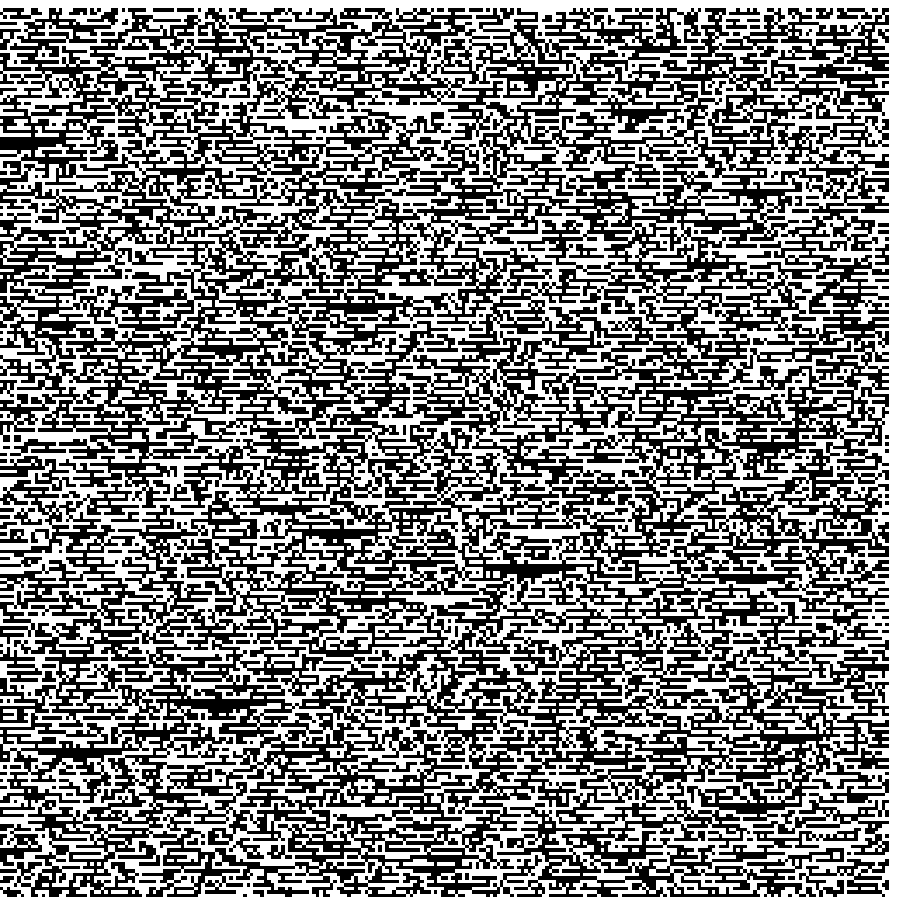} \\
\includegraphics[width=6cm]{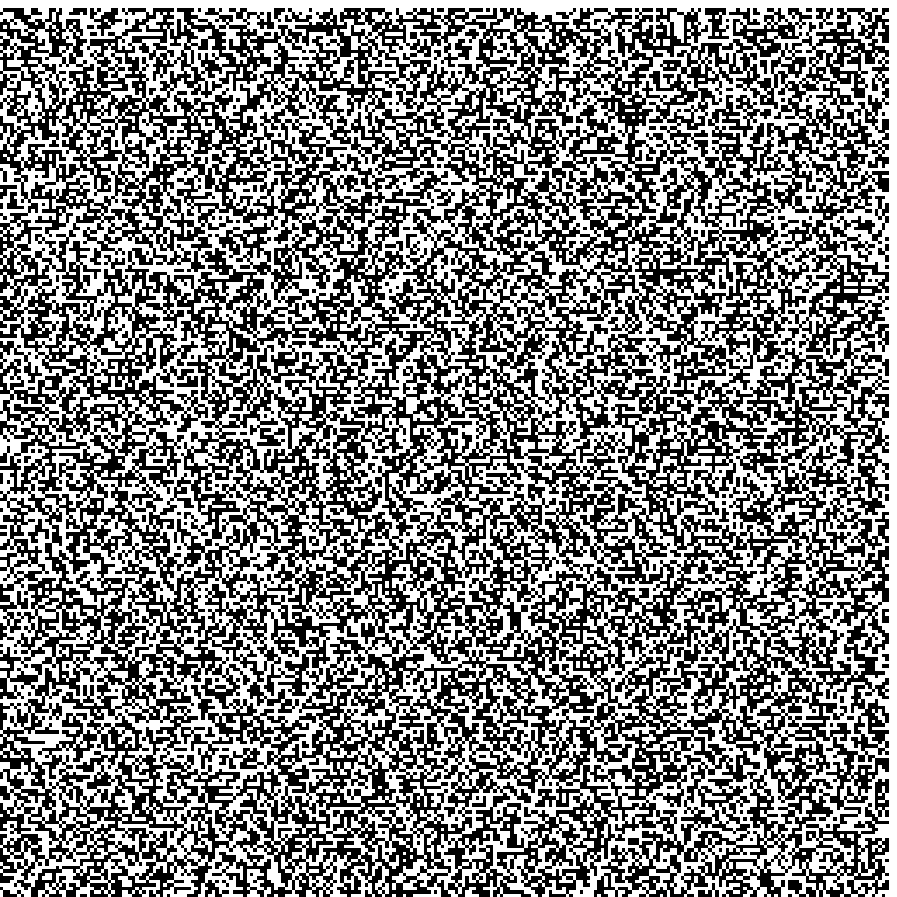} &
\includegraphics[width=6cm]{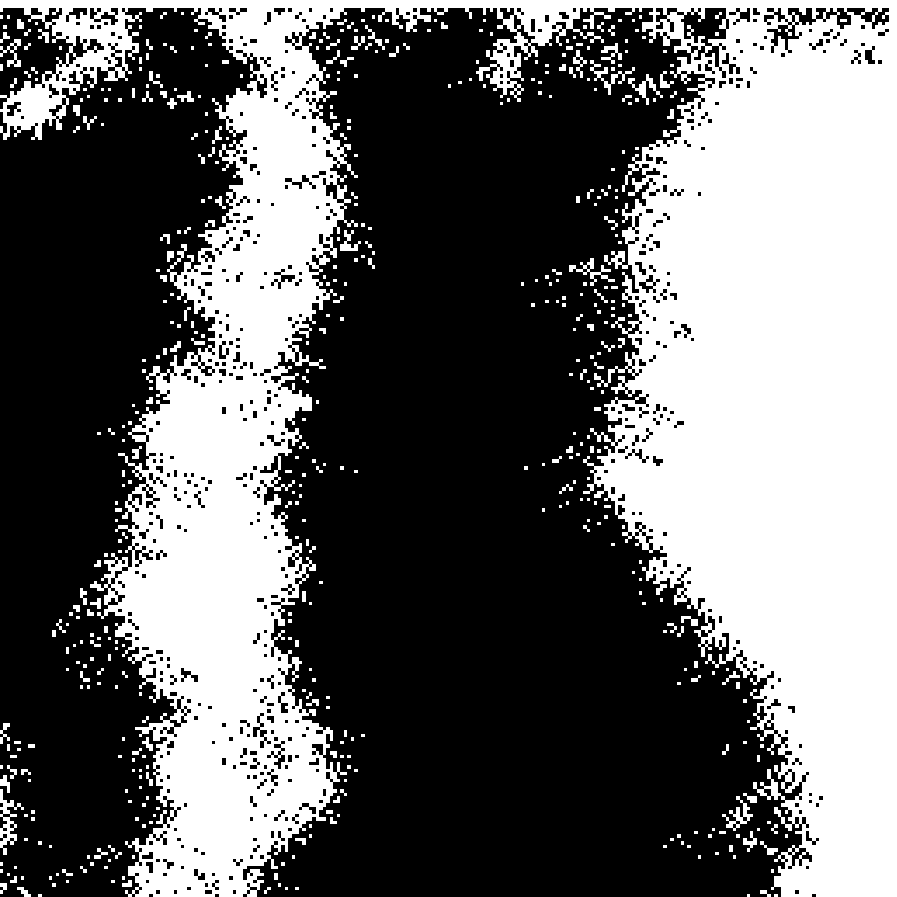}\\
\end{tabular}
\caption{\label{fig:patterns}
Typical patterns ($256\times 256$ pixels) for $R11Q1$ and $H=0$,
starting from a disordered initial condition with $\rho_0=0.5$. 
(A) [top left, $j=0.056250$]: ``chaotic'' phase. One can observe rare 
``triangles'' with ``base'' larger that $R$, 
corresponding to the unstable absorbing states, and other metastable
patterns, corresponding to additional absorbing states for
$J\rightarrow -\infty$.
(B) [top right, $j=0.421875$]: active phase, the only absorbing states
are 0 and 1. 
(C) [bottom left, $j=0.478125$]: disordered phase. 
(D) [bottom right, $j=0.562500$]: 
quiescent phase. In this phase the only stable states are the absorbing
ones. The boundaries separating the phases move randomly until coalescence.}
\end{figure}


\begin{figure}
\includegraphics[width=8cm]{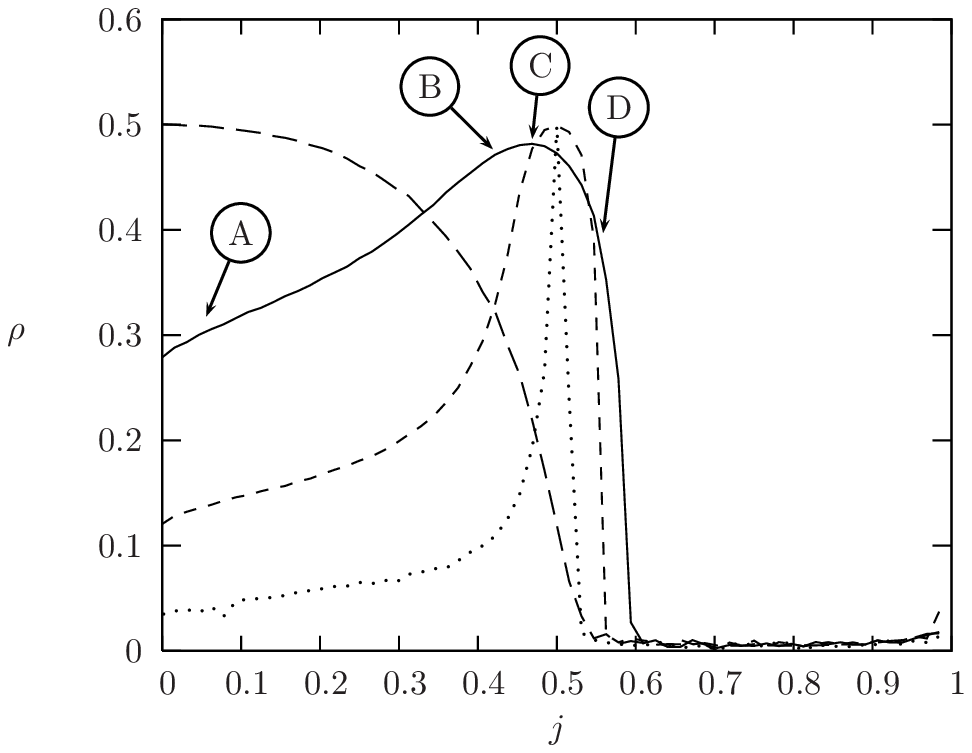} 
\caption{\label{fig:rho}Behavior of $\rho$ for $H=0$. $R3Q1$ (long dashed),
  $R11Q1$ (solid), $R41Q1$ (dashed)
and $R81Q1$ (dotted). Letters correspond to patterns in
Fig~\protect\ref{fig:patterns}.}
\end{figure}

Typical patterns for $H=0$ are shown in Fig.~\ref{fig:patterns}. Roughly speaking, for 
ferromagnetic ($J>0$) coupling, the only stable asymptotic states are the absorbing ones. The system quickly coalesce into large patches of zeroes and ones (Fig.~\ref{fig:patterns}-D), whose borders perform a sort of random motion until they annihilate pairwise. For $J<0$ the stable phase is represented by an \emph{active} state, with a mixture of zeroes and ones. For $J\rightarrow\-\infty$ the automaton become deterministic, of ``chaotic'' type (for R=3 it is Wolfram's rule 150). 

\begin{figure}
\includegraphics[width=8cm]{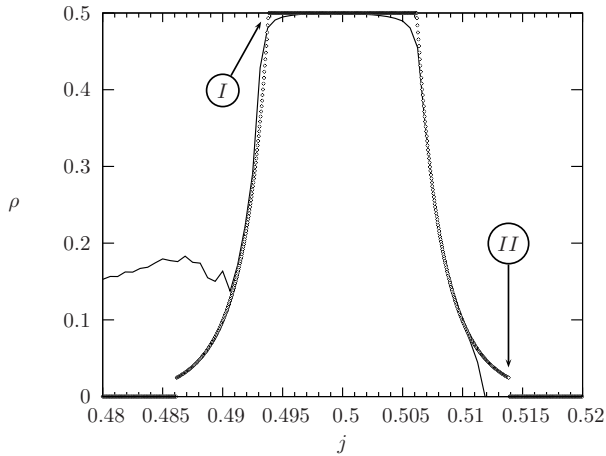}
\caption{\label{fig:H0}
Comparisons between numerical (thin line)  and mean-field (thick dotted line)
results for  $R81Q1$ and $H=0$ ($h=0.5$). The estimated critical values are $j^*_{\text{I}} \simeq 0.493827$ and $j^*_{\text{II}} \simeq 0.51384$. }
\end{figure}

As illustrated in Fig.~\ref{fig:rho}, when $R$ is greater than 3, the quantity $\rho$ is no 
more a monotonic function 
of $j$, and a new,
\emph{less disordered} phase appears inside the active one for small
values of $j$. This
phase is characterized by a large number of 
metastable states, 
that become truly absorbing only in the limit $J\rightarrow -\infty$.
For
reasons explained in the following, we shall denote the new phase by
the term \emph{irregular}, and the remaining portion of the active phase
 \emph{disordered}. By
further increasing $R$, both transitions become sharper, and the size
of the disordered phase shrinks, as shown in Fig.~\ref{fig:rho}. 
By enlarging the transition region one sees that for $H=0$
(Fig.~\ref{fig:H0}) the
transitions are composed   by two  sharp bends, which are not finite-size or time effects. As shown in Fig.~\ref{fig:patterns}-C, 
in this range of parameters the probability of observing
a local absorbing configurations (i.e.\ patches of zeroes or ones) is
vanishing. All others local configurations have finite probability of
originating zeroes or ones in the next time step. The observed
transitions are essentially equivalent to those of an equilibrium system,
that in one dimension and for short-range interactions cannot exhibit
a true phase transition. The bends thus become real salient points only
in the limit $R\rightarrow \infty$. 

The origin of the order-disorder (II) and disorder-irregular (I) phase
transition is due to the loss of stability of the fixed point $c^* \ne
0, 1$ given by the mean field approximation. More detailed considerations can be found in \cite{prl}.

For $H=0$ (Fig.~\ref{fig:H0}) the period-doubling phase brings the local
configuration into an absorbing state, and the lattice dynamics is
therefore driven by interactions among patches which are locally 
absorbing. This essentially corresponds to the dynamics of a
Deterministic Cellular Automata (DCA) of \emph{chaotic} type, i.e.\ 
a system which is insensitive of infinitesimal perturbations but
reacts in an unpredictable way to \emph{finite} perturbations.

\section{Chaos and Lyapunov exponent in cellular automata}
\label{Sec:lyap}

State variables in cellular automata are discrete, and thus the usual 
chaoticity analysis classifies them as stable systems. 
The occurrence of disordered patterns and their propagation in stable dynamical systems can be classified into two main groups: \emph{transient chaos} and \emph{stable chaos}.

Transient chaos  is an irregular behavior of finite lifetime characterized by the coexistence
in the phase space of stable attractors and chaotic non attracting sets  -- namely chaotic saddles or repellers~\cite{Tel}. After a transient irregular behavior, the system generally collapses abruptly onto a non-chaotic attractor. Cellular automata do not display this kind of behavior, which however may be present in systems having a discrete dynamics as a limiting case~\cite{stablechaos}.

Stable chaos constitutes a different kind of transient irregular
behavior \cite{CK88,PLOK} which cannot be ascribed to the presence of
chaotic saddles and therefore to divergence of nearby trajectories.
Moreover, the lifetime of this transient regime may scale
exponentially with the system size
(supertransients~\cite{CK88,PLOK}), and the final stable attractor is
practically never reached for large enough systems. One is thus
allowed to assume that such transients may be of substantial 
experimental interest and become the only physically relevant states
in the  thermodynamic limit. 

The emergence of this ``chaoticity'' in DCA dynamics, is effectively
illustrated by the damage spreading analysis~\cite{Damage1,Damage2},
which measures the sensitivity to initial conditions and for this
reason is considered as the natural extension of the Lyapunov
technique to discrete systems. In this method, indeed, one monitors
the behavior of the distance between  two replicas of the system
evolving from slightly different initial conditions. The dynamics is
considered unstable and the DCA is said chaotic, whenever a small
initial difference between replicas spreads through the whole
system.  On the contrary, if the initial difference eventually
freezes or disappears,  the DCA is considered non chaotic.  

Due to the limited number of states of the automata however, damage spreading does not account for the maximal ``production of uncertainty'', since the two replicas may synchronize locally just by chance (self-annihilation of the damage). Moreover, there are different definitions of damage spreading for the same rule~\cite{DomanyHinrichsen}.

To better understand the nature of the active phase, and up to what extent it can be denoted \emph{chaotic}, 
we extend the finite-distance Lyapunov exponent definition~\cite{LyapunovCA} to probabilistic cellular automata.
A similar approach has been used in Ref.~\cite{luque}, calculating the Lyapunov exponents of a Kauffman random boolean network in the annealed approximation. As shown in this latter paper, this computation gives the value of the (classic) Lyapunov exponent obtained by the analysis of time-series data using the Wolf algorithm.

Given a Boolean function $f(x,y,\dots)$, 
we define the Boolean derivative $\de f/\de x$, as
\[
 \dfrac{\de f}{\de x} = \begin{cases} 
  1 & \text{if $f(|x-1|, y, \dots)  \neq f(x,y,\dots)$,}\\
  0 & \text{otherwise,}
 \end{cases}
\]
which represents a measure of sensitivity of a function with respect to its arguments. The evolution rule of a probabilistic cellular automaton may be thought as a Boolean function that depends also by one or more random arguments. In our case
\[
 f(x_1, x_2\dots; r) = f(S; r) = [r<P_S],
\]
where $S=x_1+x_2,\dots$, $P_S$ is defined by Eq.~\eqref{Eq:p_s} and 
$[\text{statement}]$ is the truth function, which gives 1 if the statement is true and 0 otherwise. In this case the derivative is taken with respect to $x_i$ by keeping $r$ constant.

For a cellular automaton rule, we can thus build the Jacobian $J_{ij} = \de x_i^{t+1}/\de x_j^t$. This Jacobian depends generally on the point in phase-space (the configurations) belonging to a given trajectory. In the case of a probabilistic automaton, the trajectory also depends on the choice of the random numbers $r$. 

\begin{figure}
\includegraphics[width=8cm]{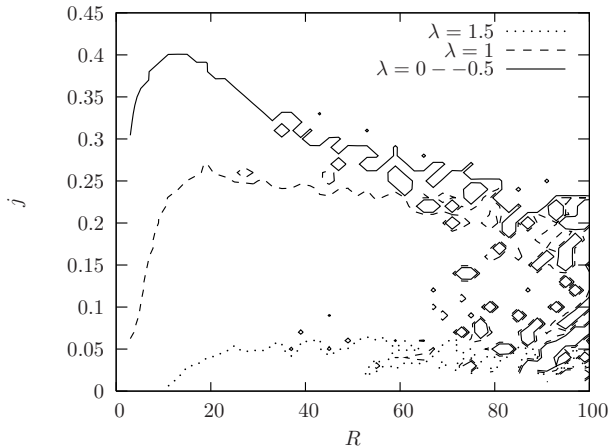}
\caption{\label{fig:Sim}
Contour plot of the maximum Lyapunov $\lambda$ exponents for different values of neighborhood size $R$ and conformist parameter $j$. The solid line represent the boundary between the $\lambda\ge 0$ phase and the $\lambda=-\infty$ one. }
\end{figure}

Finally, the maximum Lyapunov exponent $\lambda$ is computed in the usual way by
measuring the expansion rate of a ``tangent'' vector $\b{v}(t)$, whose time evolution  is given by
\[
 \b{v}(t+1) = \b{J} \b{v}(t).
\]
As explained in Ref.~\cite{LyapunovCA}, a component  $v_i(t)$ of this tangent vector may be thought as the maximum number of different paths following which any initial ($t=0$) damage may reach the site $i$ at time $t$, without self-annihilation. When all components of $\b{v}$ become zero ($\lambda=-\infty$), no information about the initial configuration may ``percolate'' to $t=\infty$, and the asymptotic configuration is determined only by the random numbers used. 
This maximum Lyapunov exponent is also related to the synchronization properties of automata~\cite{synca}. 

A preliminary numerical computation of $\lambda$
for our model is reported in Fig.~\ref{fig:Sim}. It can be noticed that the boundary $j(R)$ of $\lambda\ge0$ is not monotonic with $R$, reaching a maximum value for $R\simeq 11$. By comparisons with Fig.~\ref{fig:rho}, one can see that the chaotic phase is included in the irregular one.

\section{Conclusions}
\label{Sec:concl}

We have investigated a probabilistic cellular automaton model of an uniform society, with forcing local majority. The phase diagram of the model is extremely rich, showing a quiescent phase and several, active phases, with different dynamical behaviors. We have analyzed the properties and boundaries of these phases using direct numerical simulations, mean-field approximations and extending the notion of 
finite-distance Lyapunov exponent to probabilistic cellular automata.

\end{document}